\documentclass{elsarticle}
\usepackage{graphicx}
\usepackage{siunitx}
\usepackage{hyperref}
\usepackage{float}
\usepackage[T1]{fontenc}
\usepackage[utf8]{inputenc}

\begin{document}

\begin{frontmatter}

  \title{Generative Artificial Intelligence for Air Shower Simulation}

  \author[a,b]{C. Bozza}
  \author[a,b]{A. Calivà}
  \author[a,b]{A. De Caro}
  \author[a,b]{D. De Gruttola}
  \author[a,b]{S. De Pasquale}
  \author[a,b]{L.A. Fusco}
  \author[a,b]{G. Messuti\corref{cor1}}
  \author[a,b]{C. Poirè}
  \author[a,b]{S. Scarpetta}
  \author[a,b]{T. Virgili}

  \address[a]{Dipartimento di Fisica ``E.R. Caianiello'', Università degli Studi di Salerno,
    Via Giovanni Paolo II, 132, 84084 Fisciano (SA), Italy}

  \address[b]{Istituto Nazionale di Fisica Nucleare, Gruppo Collegato di Salerno,
    at Università degli Studi di Salerno,
    Via Giovanni Paolo II, 132, 84084 Fisciano (SA), Italy}

  \cortext[cor1]{Corresponding author: \href{mailto:gmessuti@unisa.it}{gmessuti@unisa.it}}

  \begin{abstract}    
    The detailed simulation of extensive air showers, produced by primary cosmic rays interacting in the atmosphere, is a task that is traditionally undertaken by means of Monte Carlo methods. These processes are computationally intensive, accounting for a major fraction of the computational resources used in the large-scale simulations required by current and future experiments in the field of astroparticle physics. In this work, we present a novel approach based on Generative Adversarial Networks (GANs) to accelerate air shower simulations. We developed and trained a GAN on a dataset of high-energy proton-induced air showers generated with \texttt{CORSIKA}; our model reproduces key distributions of secondary particles, such as energy spectra and spatial distributions at ground level of muons. Once the model has been trained, which takes approximately 74 hours, the generation real time per shower is reduced by a factor of $10^4$ with respect to the full \texttt{CORSIKA} simulation, leading to a substantial decrease in both computational time and energy consumption.
  \end{abstract}

  \begin{keyword}
    Cosmic Rays, Air Shower Simulation, Generative Adversarial Networks, High-Performance Computing, Machine Learning
  \end{keyword}

\end{frontmatter}

\section{Introduction} \label{sec:intro}

High-energy ($E\gg\si{\tera\electronvolt}$) cosmic rays are messengers from some of the most extreme environments of the Universe. These energies are indeed clearly not achievable by means of thermal mechanisms in astrophysical environments, and require the presence of shock waves or relativistic jets propagating through the interstellar medium and radiation fields at or near cosmic accelerators~\cite{bib:spurio}.

When \textit{primary} cosmic rays enter the Earth's atmosphere, they interact via the strong nuclear force with atomic nuclei, and initiate cascades of \textit{secondary} particles, Extensive Air Showers (EAS)~\cite{Kampert:2012}. The terrestrial atmosphere acts as a calorimeter, and can thus be used to test the properties of the incoming cosmic rays by observing these showers with ground-based particle detectors, Imaging Air-Cherenkov Telescopes, or large-volume neutrino telescopes located deep under water or under ice. Such observations allow the study the properties of the primary cosmic ray flux (its intensity, energy spectrum, and chemical composition) as well as the fundamental hadronic physics at energies beyond the reach of terrestrial accelerators~\cite{PDG:CosmicRays}.

The accurate interpretation of experimental data relies on detailed simulations of the EAS development. In these simulations, typically carried out by means of Monte Carlo (MC) methods, the primary cosmic ray interaction is simulated, and its products are transported through the atmosphere up to the point where they either interact or decay. Consequently, the particles emerging from such interactions or decays are also tracked downstream, until the shower dies out due to energy loss processes, or because these particles reach ground level or the detector elements. To mitigate the high computational cost of these detailed full MC simulations, faster parametric generators based on analytical descriptions or hybrid approaches have been developed~\cite{Bergmann:2007}. For the highest fidelity and detail, the most common solution in astroparticle physics is the \texttt{CORSIKA} (COsmic Ray SImulations for KAscade) MC software~\cite{Heck:1998}. While highly accurate, this full-simulation approach is demanding in terms of computational resources. Indeed, the simulation time of the full EAS scales linearly with the number of particles to be tracked, which can be demonstrated to be, in a simple approximation~\cite{heitler}, directly proportional to the energy of the primary cosmic ray. As such, the simulation time for a single high-energy EAS can range from a few seconds/minutes (in the \SIrange{1}{100}{\tera\electronvolt} energy range) to hours/days (in the \si{\exa\electronvolt} range) on modern CPU cores; since large statistical samples of these high-energy and ultra-high-energy events are needed in current and next-generation cosmic-ray experiments, this often represents a large fraction of the total computational effort required by experimental Collaborations. In addition, these large CPU requirements also carry a significant energy footprint if resources are not carefully managed.

The computational challenge sketched above hence motivates the exploration of data-driven approaches, where the CPU-intensive generation phase is replaced by rapid inference from a trained model. Among these techniques, Generative Adversarial Networks (GANs)~\cite{Goodfellow:2014} offer a promising alternative to MC simulations. GANs have demonstrated remarkable success in generating complex, high-dimensional data in various fields and have recently been explored for applications in high-energy physics, such as shower simulation in calorimeters~\cite{Paganini:2018}. In this paper, we present a GAN-based model (\texttt{GAIAS2} --- Generative Artificial Intelligence for Air Shower Simulation) that has been designed to generate distributions of muons at ground level from proton-induced air showers. We here focus on the muon component as it is a crucial observable for discriminating the mass of the primary cosmic ray, and also constitutes a primary background for underwater and under-ice neutrino observatories. 

\section{Monte Carlo simulations} \label{sec:mcsim}

The datasets used to train and evaluate our generative model were produced with the \texttt{CORSIKA} software, version 7.7410~\cite{Heck:1998} on a computing cluster dedicated to this task. The Monte Carlo samples used here assume a planar detector surface located at ground level (at sea level), below a ``standard'' atmosphere, as defined in \texttt{CORSIKA}.

Since our primary goal was to generate a large dataset that would be suitable for training a generative model, we focused on showers initiated by primary protons. \texttt{CORSIKA} simulations were run so that a wide energy range, from \SI{1}{\tera\electronvolt} to \SI{300}{\peta\electronvolt}, would be covered in the GAN training. To ensure uniform statistics across the whole range, events were generated according to a power-law energy spectrum with a spectral index of $\gamma = -1$, which produces a flat differential distribution in the logarithm of the primary energy. The arrival direction of the primary protons was chosen to be the local vertical, to feed the training phase with a homogeneous sample of air showers. Differences in the first interaction point of the simulated primary protons would yield differences in the depth development of the shower in the atmosphere, so that the training sample could be composed of a sufficiently varied set of simulated EAS. The coordinate system for these events is centred along the extensive air shower axis which, because of momentum conservation, tends to correspond to the incoming direction of the proton.

The development of the air shower in the atmosphere is critically dependent on the description of the hadronic interactions occurring as particles propagate in the atmosphere. To account for this, we generated four different datasets in which different combinations of two high-energy and two low-energy physics models were used. For high-energy interactions, the two hadronic interaction models (HIM) considered were EPOS-LHC~\cite{Pierog:2015} and SIBYLL 2.3c~\cite{bib:sib23c}. At low energies, these were paired with either the UrQMD~\cite{UrQMD} or GEISHA~\cite{geisha} HIM. Even though updated models are available~\cite{bib:epos_lhcr,Riehn:2020}, especially in light of more recent measurements at particle accelerators and of further tuning of the HIMs to resolve data/Monte Carlo discrepancies in CR data (e.g, the ``muon puzzle''~\cite{muon_puzzle}), these proof-of-concept simulations for the training of the GAN are still valid.

\begin{figure}[htb]
  \centering
  \begin{minipage}{0.45\textwidth}
    \includegraphics[width=\textwidth]{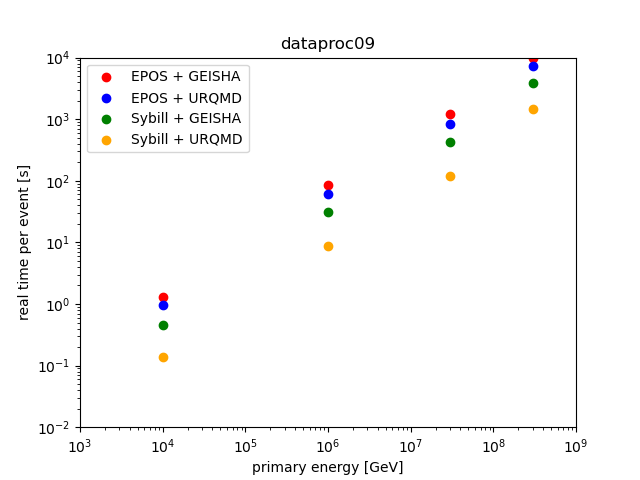}
  \end{minipage}
  \hspace{0.05\textwidth}
  \begin{minipage}{0.45\textwidth}
    \includegraphics[width=\textwidth]{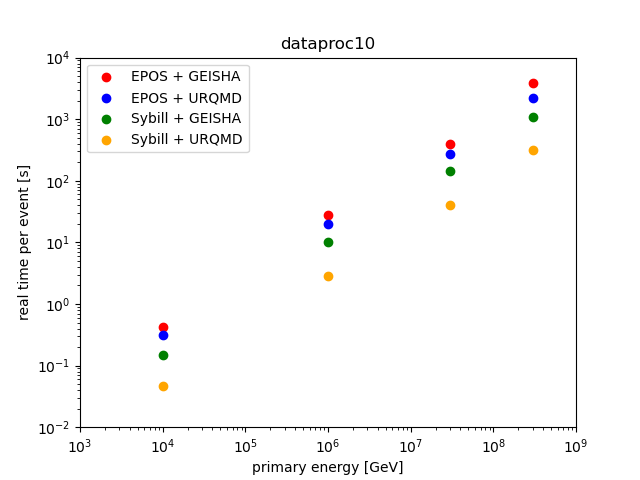}
  \end{minipage}\\
  \begin{minipage}{0.45\textwidth}
    \includegraphics[width=\textwidth]{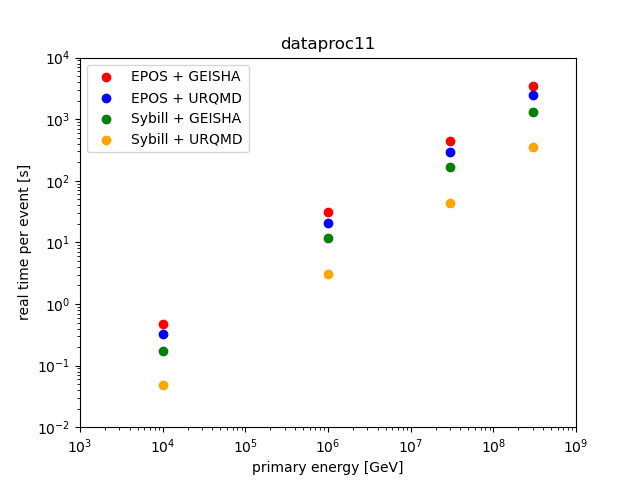}
  \end{minipage}
  \hspace{0.05\textwidth}
  \begin{minipage}{0.45\textwidth}
    \includegraphics[width=\textwidth]{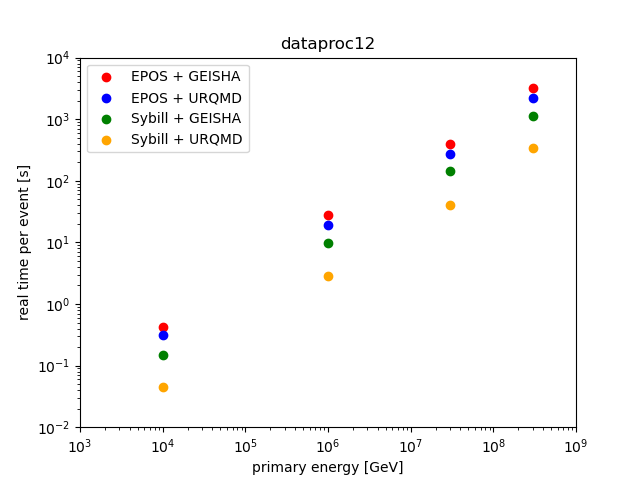}
  \end{minipage}
  \caption{Processing time per event (real time, as from the Unix command \texttt{time}) for benchmark simulations in the four different servers used in this work for the Monte Carlo generation of the training samples. The different colours reported in the figures account for the different hadronic interaction models used in the simulations (see text for more details).}
  \label{fig:cputimes}
\end{figure}

The computing cluster on which the Monte Carlo simulations were run is composed of four servers, named here \texttt{dataproc09-10-11-12}. The CPUs used for the Monte Carlo production are listed in Tab.~\ref{tab:cpus}. One of the servers (\texttt{dataproc10}) was benchmarked using the \texttt{HEPSCORE23} version 2 procedures~\cite{hepspec}, which returned a CPU time scaling factor of 10 (1 hour runtime on \texttt{dataproc10} is equivalent to 10 standardised CPU hours). The computing time of the other servers could then be rescaled to standardised CPU hours by comparing the processing time of equivalent Monte Carlo simulations that were reprocessed with exactly the same inputs. The \texttt{Slurm} workload manager~\cite{slurm} was used to schedule the Monte Carlo production jobs. The simulation time (real time, as returned by the Unix \texttt{time} command) for benchmark simulations on different servers is reported in Fig.~\ref{fig:cputimes}. 

\begin{table}[htb]
  \centering
  \begin{tabular}{l|l}
    Server ID & CPU \\
    \hline
    \texttt{dataproc09} & AMD Opteron(tm) Processor 6168 $@$ \SI{1.9}{\giga\hertz} \\
    \texttt{dataproc10} & $2\times$ Intel(R) Xeon(R) Silver 4316 CPU $@$ \SI{2.30}{\giga\hertz} \\
    \texttt{dataproc11} & $2\times$ Intel(R) Xeon(R) Gold 6252N CPU $@$ \SI{2.30}{\giga\hertz} \\
    \texttt{dataproc12} & $2\times$ Intel(R) Xeon(R) Silver 4316 CPU $@$ \SI{2.30}{\giga\hertz}
  \end{tabular}
  \caption{List of the CPUs used to produce the Monte Carlo simulations exploited in the training of \texttt{GAIAS2}.}
  \label{tab:cpus}
\end{table}

The final high-statistics production consists of 2 million showers, split across 10 distinct energy bins to optimise CPU time usage. This full Monte Carlo production consumed approximately \num{5000000} standardised core-hours and resulted in 30 TB of stored data. To ensure reproducibility, the containerised environments and automation scripts are publicly available on \texttt{github}~\cite{GAIAS2:Github}. From each simulated event, the state vectors of all muons reaching the observation level with an energy larger than \SI{1}{\giga\electronvolt} at sea level were stored, including their energy, position relative to the shower core, and momentum components. This muon dataset constitutes the truth for the training and validation of the GAIAS2 model described in the following section.

While the \texttt{CORSIKA} software is currently being rewritten and updated to its version 8 (\texttt{CORSIKA8}~\cite{Huege:2023}), the version 7 of the software is still the standard in the astroparticle physics community, and thus we will only focus on its output as the training sample for our GAN. In any case, following reports found in the literature~\cite{corsika8_comp}, we also tested MC productions using version 8 of the software. This test was run considering the monochromatic generation of vertical protons in the same standard atmosphere, for the same planar detector geometry at ground level, and for the same set of corresponding HIM (barring slight differences in the HIM versions, e.g., for the Sibyll model). Then, we compared the longitudinal and lateral distributions of secondary particles at ground level and at the EAS maximum under these hypotheses. All tests performed are in agreement with the literature, with differences in the range of \SIrange{1}{10}{\percent}. It should be noted that \texttt{CORSIKA8} is still approximately \SI{10}{\percent} slower than version 7, given that the latter software has had a long history of optimisation as also reported in~\cite{corsika8_comp}.

\subsection{Benchmark observables for MC simulations}

\begin{figure}[hbt]
  \begin{center}
    \begin{minipage}{0.8\textwidth}
      \includegraphics[width=\textwidth]{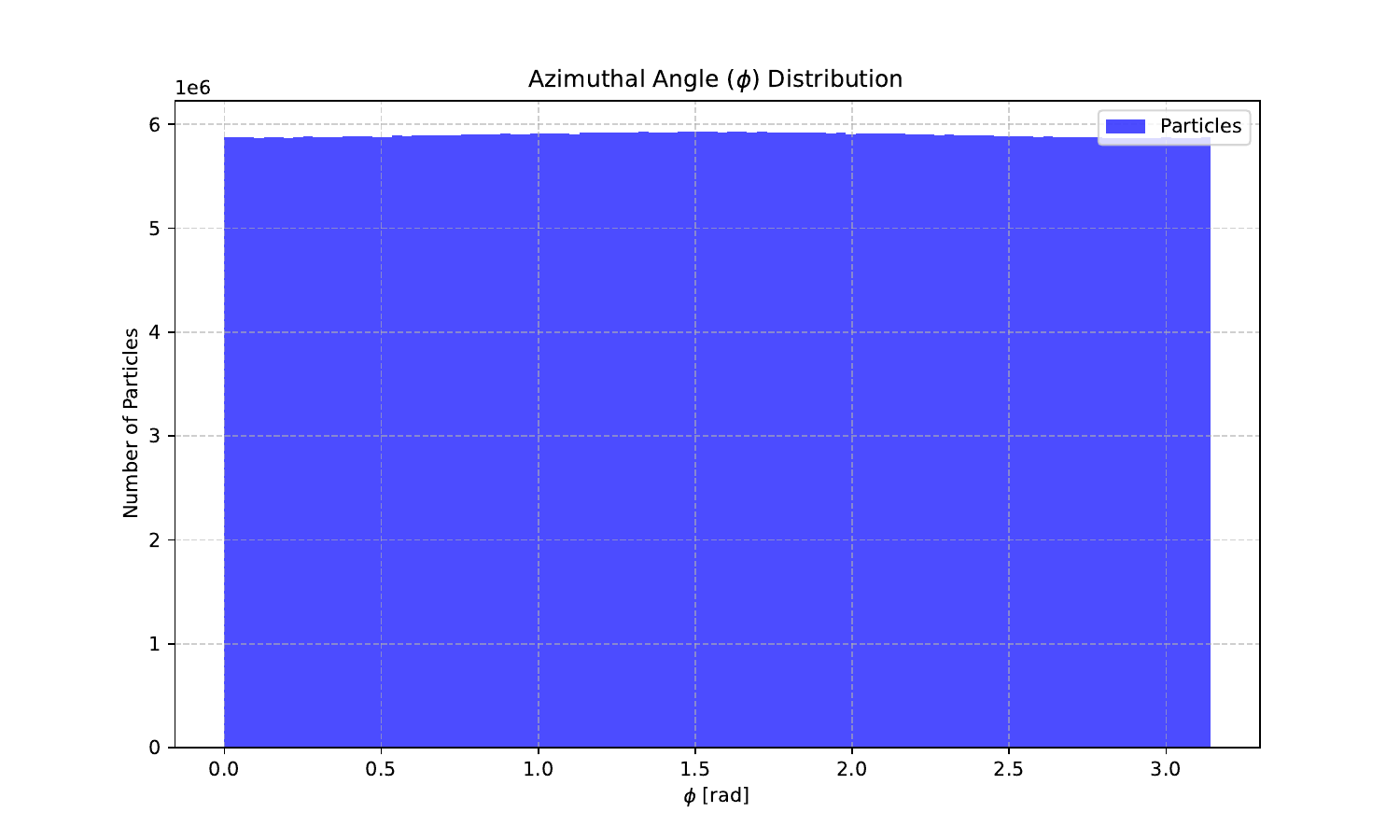}
    \end{minipage}\\
    \begin{minipage}{0.8\textwidth}
      \includegraphics[width=\textwidth]{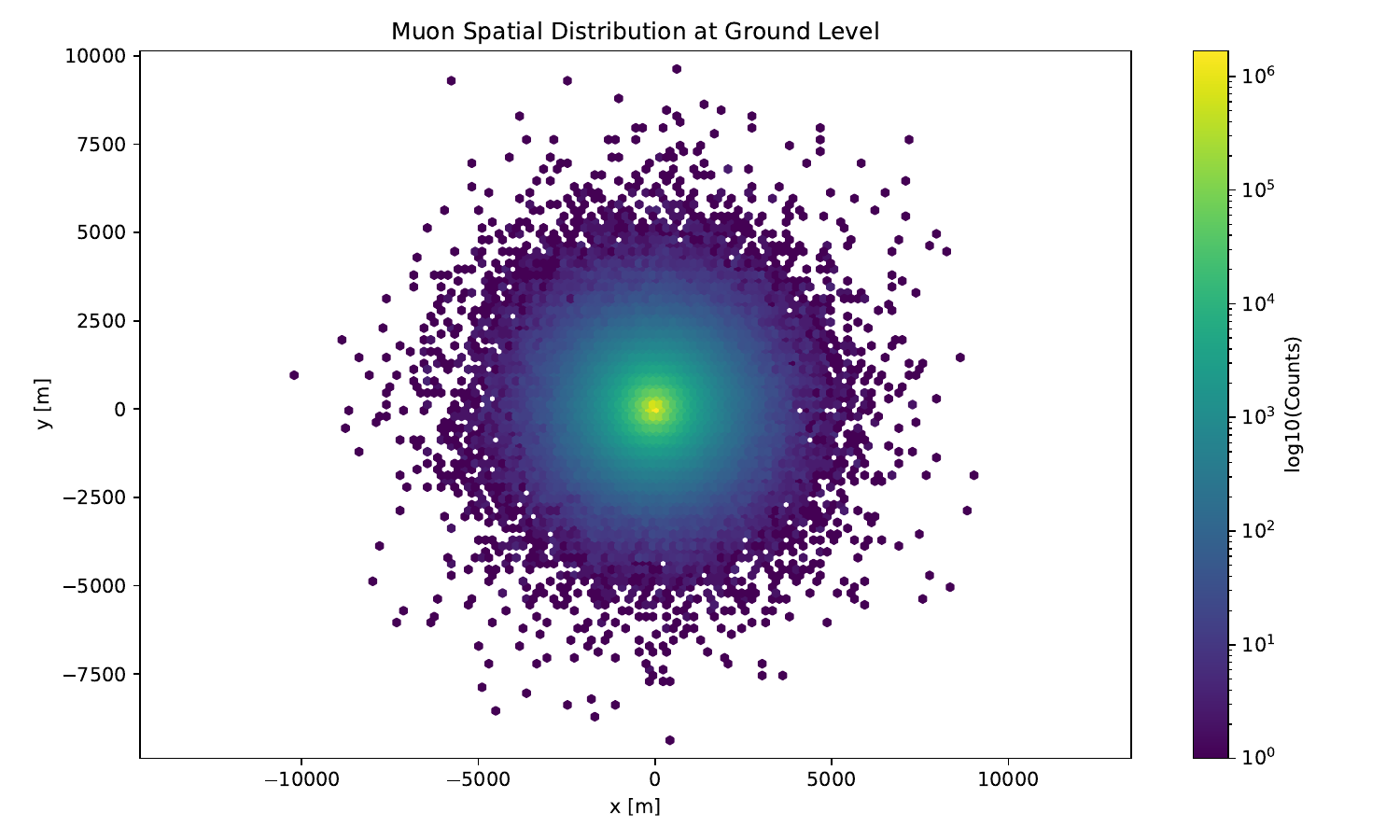}
    \end{minipage}
  \end{center}
  \caption{Azimuthal angle ($\phi$) distribution (top) and spatial distribution of muon positions in the $(x, y)$ plane (bottom) at ground level.}
  \label{fig:cylindrical_symmetry}
\end{figure}

To validate the MC simulations produced with \texttt{CORSIKA}, several benchmark observables were analysed. Since the primary cosmic proton is injected vertically into the atmosphere, the resulting air shower is expected to exhibit cylindrical symmetry about the shower axis. This symmetry was verified from the azimuthal distribution of secondary particles, and from the two-dimensional spatial distribution of muons at ground level. As shown in Fig.~\ref{fig:cylindrical_symmetry}, the azimuthal angle ($\phi$) distribution is uniform, and the $(x, y)$ positions of muons are symmetric with respect to the shower axis.  At ground level, the radial extension of air showers initiated by primary protons with energies in the range \SIrange{e6}{e7}{\giga\electronvolt} extends up to approximately \SI{10}{\kilo\metre}.

\begin{figure}[hbt]
  \begin{center}
    \includegraphics[width=0.8\textwidth]{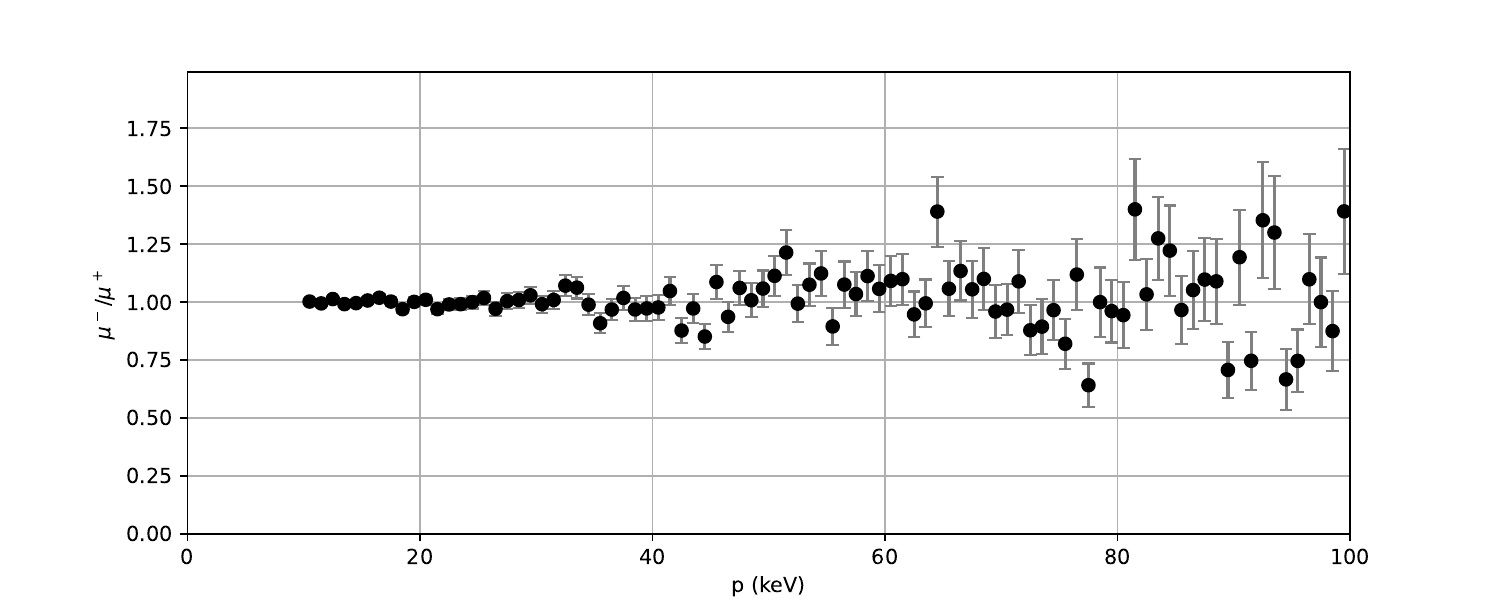}
  \end{center}
  \caption{Ratio of negative to positive muons ($\mu^-/\mu^+$) at ground level as a function of momentum.}
  \label{fig:ratio_muons}
\end{figure}

The ratio of negative to positive muons ($\mu^-/\mu^+$) at ground level is a particularly useful benchmark observable in air-shower simulations. Since most muons are produced from the decay of charged pions and kaons in the shower, this ratio provides a sensitive test of the underlying hadronic interaction models implemented in \texttt{CORSIKA}. The ratio of negative to positive muons at ground level, shown in Fig.~\ref{fig:ratio_muons} as a function of momentum, is found to be consistent with unity within the statistical uncertainties. This result indicates that the expected charge asymmetry, originating from the positive charge of the primary cosmic protons, is small at ground level, largely washed out during the development of the air shower.

\begin{figure}[hbt]
  \begin{center}
    \begin{minipage}{0.8\textwidth}
      \includegraphics[width=\textwidth]{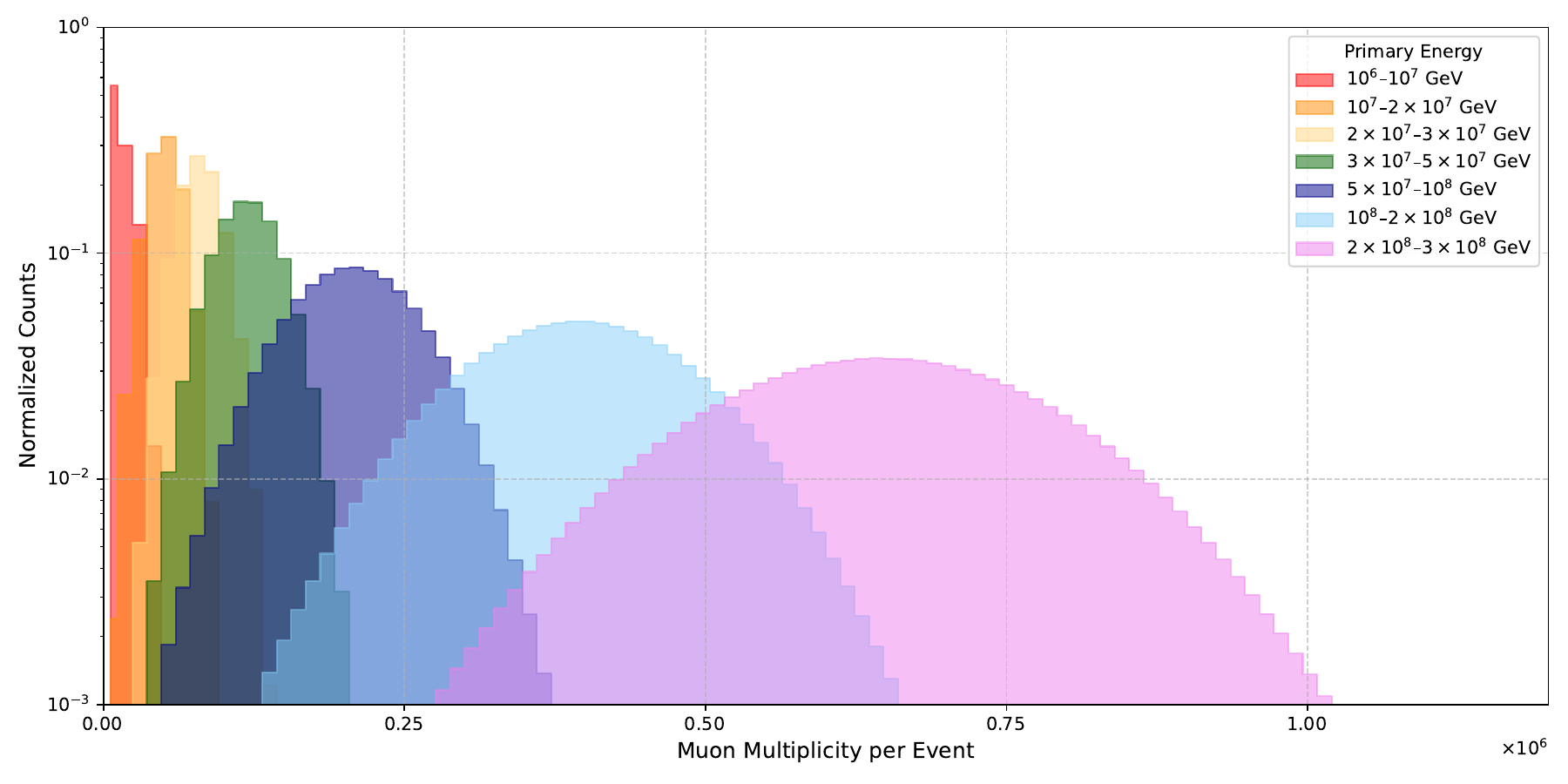}
    \end{minipage}\\
    \begin{minipage}{0.8\textwidth}
      \includegraphics[width=\textwidth]{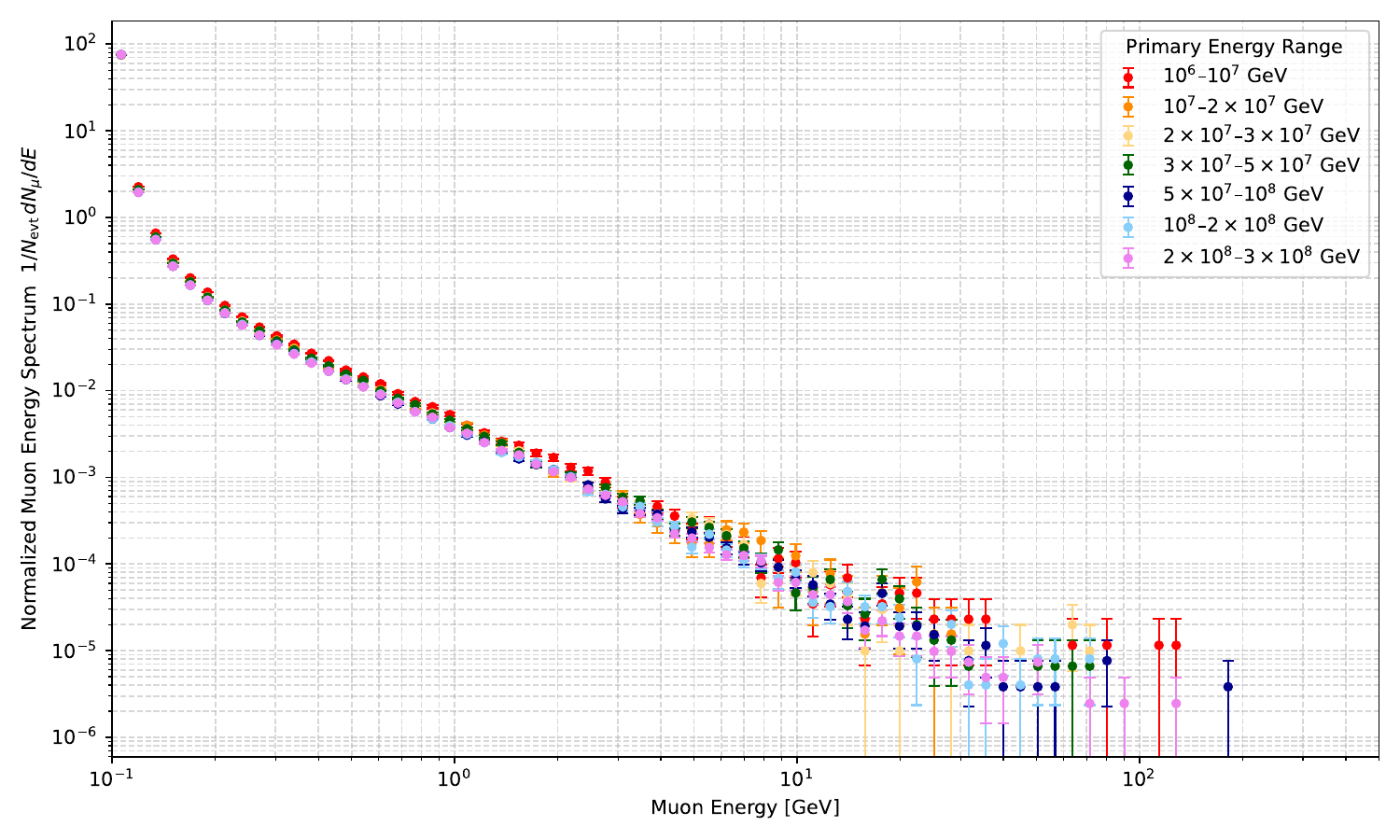}
    \end{minipage}
  \end{center}
  \caption{Muon multiplicity distributions at ground level (top) and normalised energy spectra (bottom) for air showers induced by primary protons of different energies.}
  \label{fig:muon_spectra}
\end{figure}

The muon energy spectra were studied for different primary energy intervals. While the total muon yield increases with the primary energy, as illustrated in Fig.~\ref{fig:muon_spectra} (top), the spectral shape remains nearly unchanged, as seen in the normalised distributions in Fig.~\ref{fig:muon_spectra} (bottom). This result indicates that the primary energy is shared among an increasing number of secondary particles, with each particle receiving a similar fraction of energy, resulting in a similar spectral shape at ground level.

\begin{figure}[hbt]
  \begin{center}
    \includegraphics[width=0.8\textwidth]{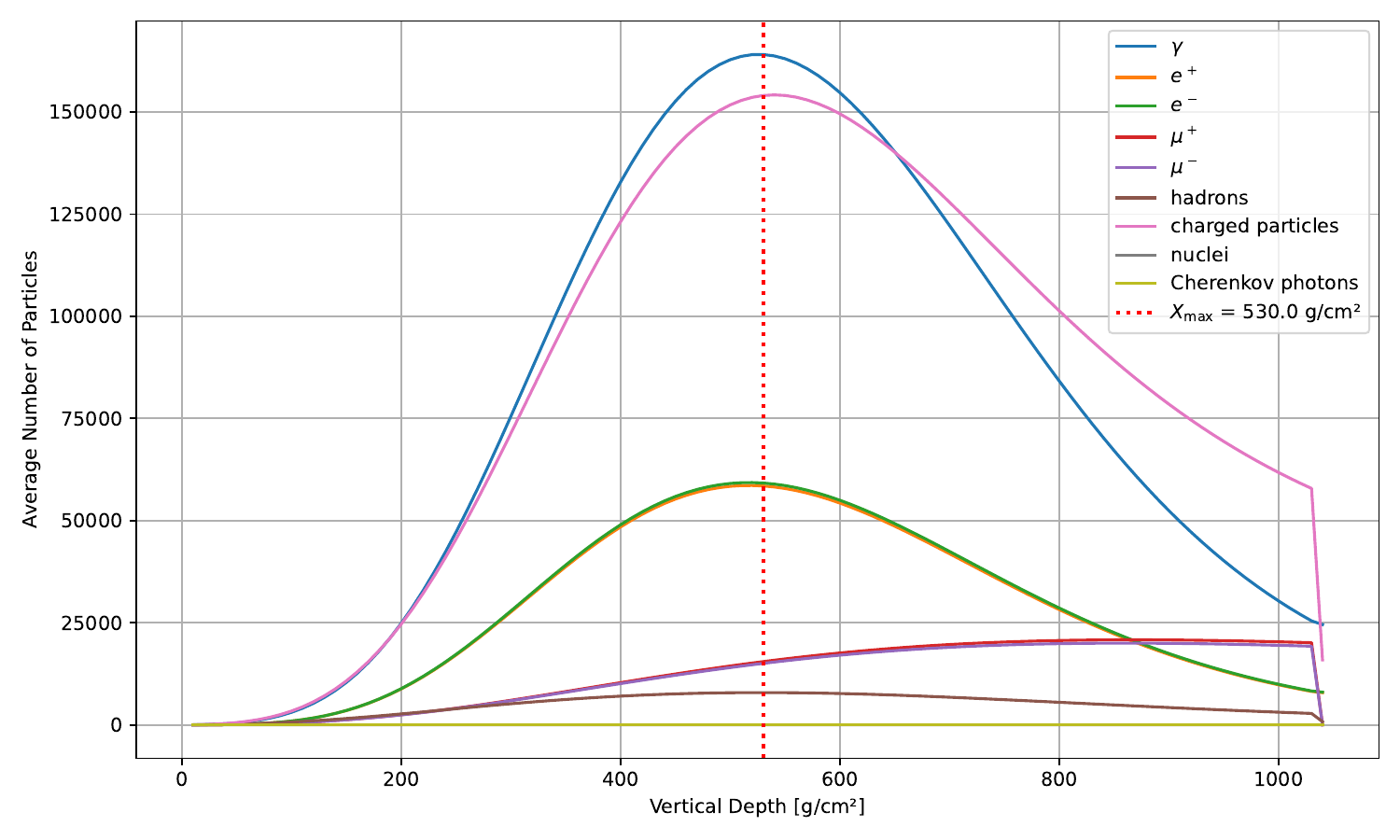}
  \end{center}
  \caption{Longitudinal profile of the air shower. The vertical red dotted line marks the atmospheric depth corresponding to the shower maximum.}
  \label{fig:longitudinal_profile}
\end{figure}

The longitudinal development of the air shower was characterised by studying the average number of particles of different species as a function of the vertical atmospheric depth (Fig.~\ref{fig:longitudinal_profile}). The depth corresponding to the shower maximum, indicated by the vertical dotted line, is found to be around \SI{530}{\gram\per\centi\metre\squared} for primary protons with energies between \num{e6} and \SI{e7}{\giga\electronvolt}.

\begin{figure}[hbt]
  \begin{center}
    \begin{minipage}{0.8\textwidth}
      \includegraphics[width=\textwidth]{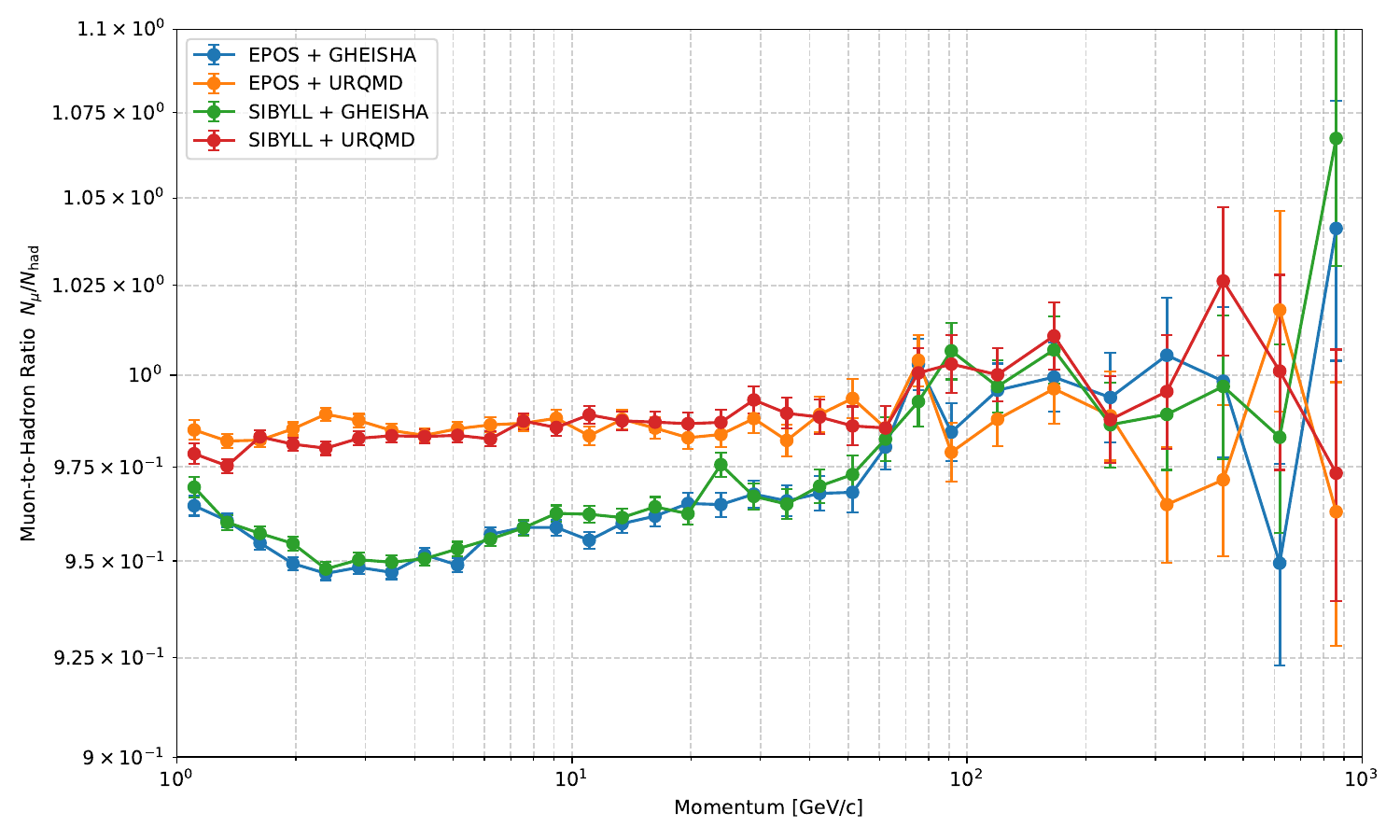}
    \end{minipage}\\
    \begin{minipage}{0.8\textwidth}
      \includegraphics[width=\textwidth]{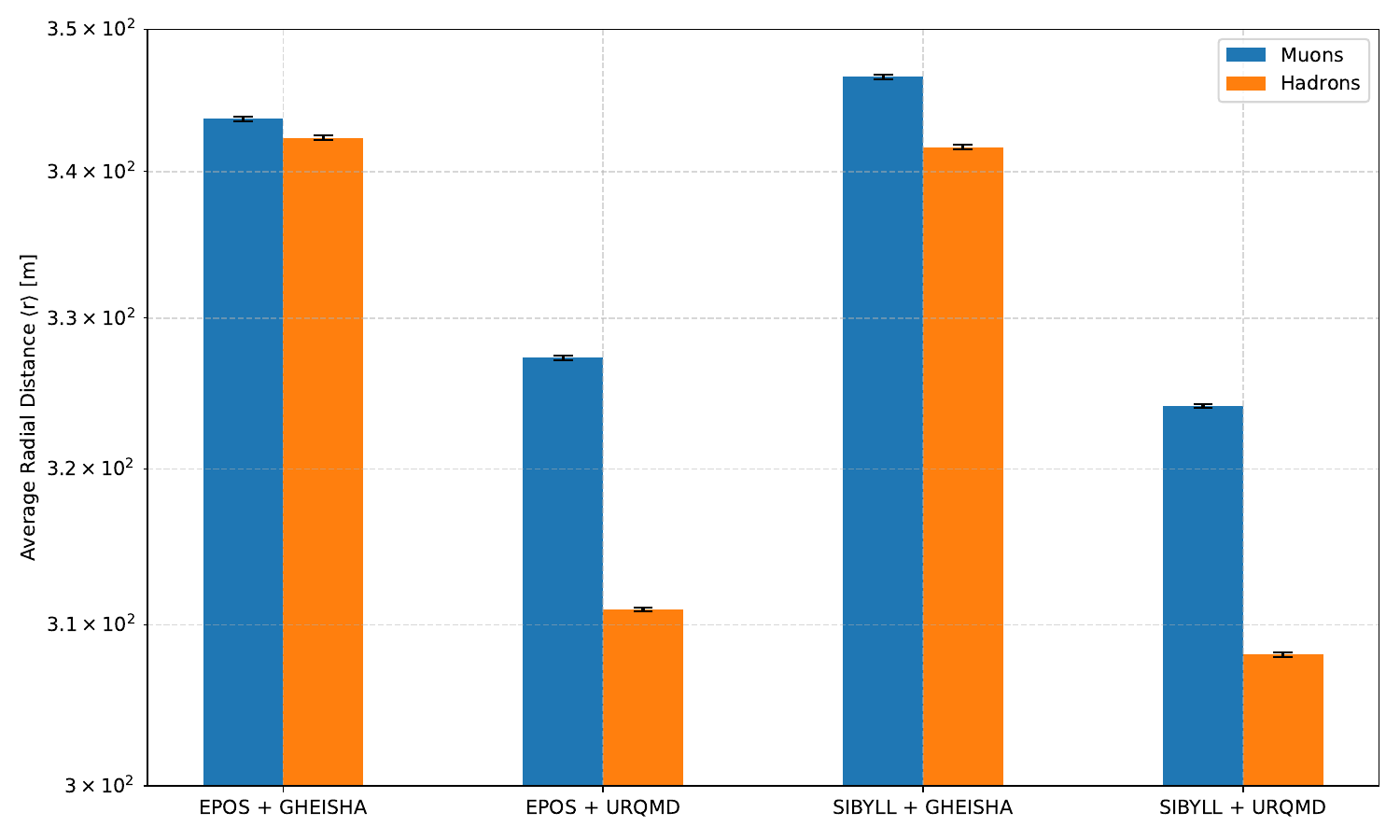}
    \end{minipage}
  \end{center}
  \caption{Muon-to-hadron ratio ($\mu/h$) at ground level (top) and average radial extension of the shower (bottom) for different hadronic interaction models.}
  \label{fig:different_hadronic_models}
\end{figure}

Differences among hadronic interaction models were investigated through the muon-to-hadron ratio ($\mu/h$) (Fig.~\ref{fig:different_hadronic_models}, top), and the average radial extension of the shower at ground level  (Fig.~\ref{fig:different_hadronic_models}, bottom). A maximum deviation of about \SI{2.5}{\percent} is observed in the $\mu/h$ ratio for momenta below \SI{100}{\giga\electronvolt}, whereas at higher momenta the models are consistent within uncertainties. The average radial extension differs by up to about \SI{11}{\percent} between models, indicating modest variation in the spatial development of the cascade.

\section{Data representation and preprocessing} \label{sec:Data_preprocess}

Convolutional neural networks~\cite{LeCun:2015} operate on fixed-size tensors with well-defined boundaries and binning schemes. To adapt the data produced with \texttt{CORSIKA} to the convolutional structure of our generative network, the raw particle list for each shower event was converted into a fixed-size, four-dimensional tensor. This tensor serves as a histogram of the muon distribution in a 4D phase space defined by the momentum components ($p_x, p_y, p_z$) and the transverse distance from the shower axis ($r = \sqrt{x^2 + y^2}$). The final tensor has a shape of $(16 \times 16 \times 16 \times 8)$. Each cell in this tensor contains logarithmically scaled particle counts, given by $\frac{\log_{10}(N+1)}{M}$, where $N$ is the number of muons falling within that specific bin and $M$ is a normalisation constant introduced to confine the resulting values to the interval $[0,1]$.

\begin{figure}[hbt]
  \centering
  \includegraphics[width=0.75\textwidth]{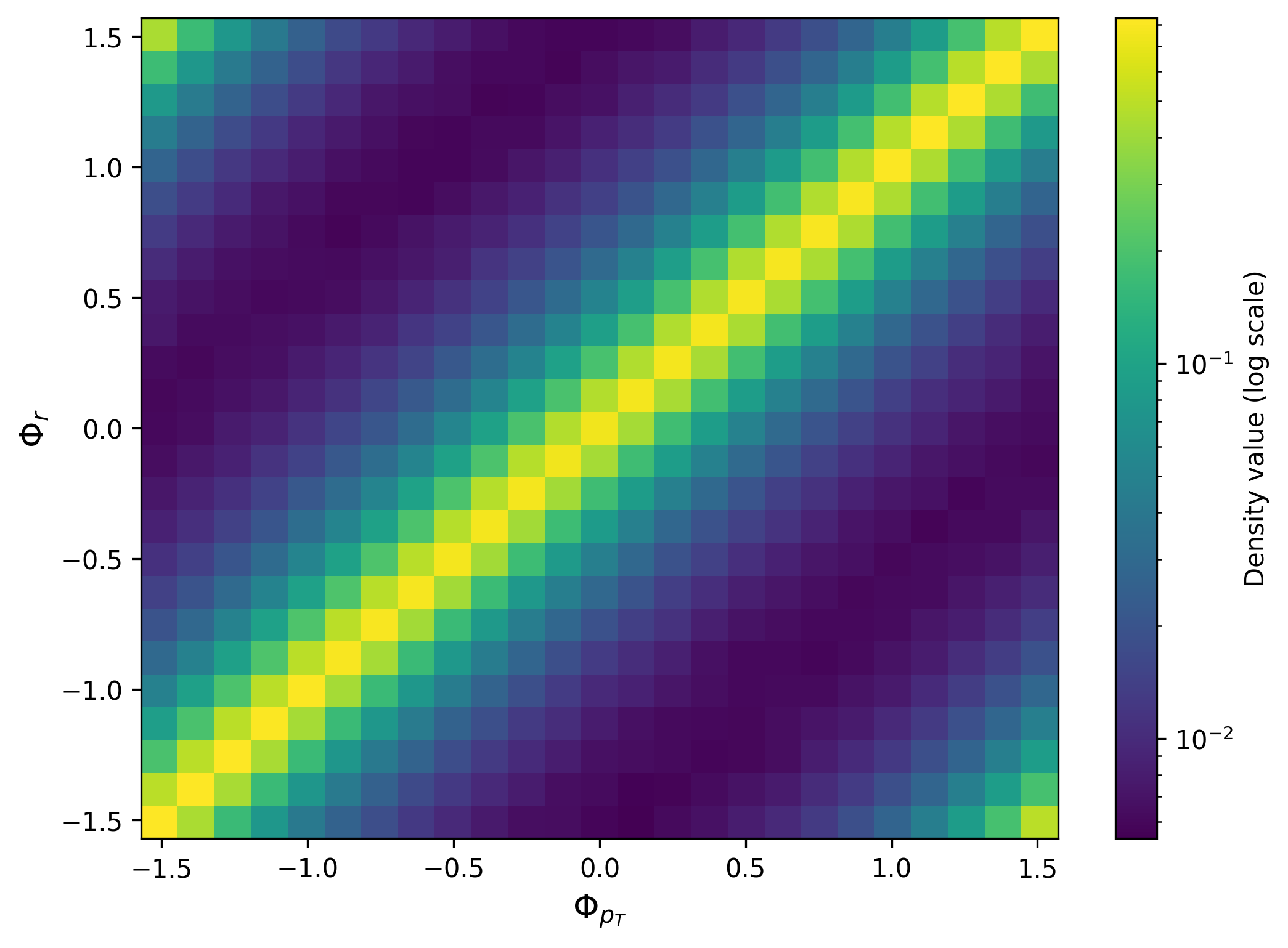}
  \caption{Two-dimensional density histogram comparing the azimuthal angle of the particle position ($\phi_r$) and the transverse momentum direction ($\phi_{p_T}$) for all simulated events with primary energies in the range \SIrange{e6}{2e7}{\giga\electronvolt}. The two angles exhibit a circular correlation value of $\rho_c = 0.94$.}
  \label{fig:Pt_vs_rt}
\end{figure}

The reduction of the transverse spatial coordinates $(x, y)$ to the single radial variable $r$ is based on the simplifying assumption, validated using our simulation data (see Fig.~\ref{fig:Pt_vs_rt}), that the azimuthal angle of a particle at sea level is strongly correlated with the angle of its transverse momentum vector, $\vec{p}_T$. This is physically motivated, since high-energy particles tend to travel in a straight line from their production point high in the atmosphere. 

The two-dimensional distribution of the azimuthal angle of the particle position $\phi_r$ versus that of its transverse momentum $\phi_{p_T}$ is shown in Fig.~\ref{fig:Pt_vs_rt}. The data points are concentrated along the diagonal (and around the upper-left and bottom-right corners), indicating a strong one-to-one correspondence between the two quantities. To quantify this relationship, we computed the circular correlation coefficient $\rho_c$, whose values lie in the interval $[-1,1]$ and measured the correlation between angular variables while accounting for their periodic nature ~\cite{Jammalamadaka:2001}. A value of $\rho_c = 0.94$ was found, indicating a strong correlation.

\subsection{Data-driven binning scheme}

A data-driven binning scheme is implemented to ensure that the fixed-size tensor representation efficiently captures the most populated regions of the phase space. 

The bin boundaries for each of the four variables are determined as follows:
\begin{itemize}
\item \textbf{Transverse momenta} ($p_x, p_y$): the 16 bins for each component are linearly spaced within a symmetric interval $[-P_M, +P_M]$. The maximum value $P_M$ is set to the 80$^{\text{th}}$ percentile of the distribution of maximum absolute momenta per event, effectively excluding rare, high-$p_T$ outliers.
\item \textbf{Transverse distance} ($r$): the 8 bins are logarithmically spaced up to a maximum radius $R_M$, defined by the 90$^{\text{th}}$ percentile of the maximum radial distances per event. The first bin includes particles whose $r$ values range from 0 up to the first logarithmic boundary. 
\item \textbf{Longitudinal momentum} ($p_z$): 16 bins are constructed. The binning procedure starts from 32 logarithmically spaced intervals covering the range from \SI{1}{\giga\electronvolt} up to the maximum observed $p_z$ value in the training dataset. The first eight bins are merged in pairs (to reduce the overabundant resolution at low energies), resulting in 4 bins; the following 11 are kept unchanged; the last thirteen are combined into a single wider bin to avoid sparsely populated regions at high momenta. This adaptive adjustment ensures a more uniform bin occupancy across the full $p_z$ range.
\end{itemize}

\begin{figure}[hbt]
  \centering
  \includegraphics[width=\textwidth]{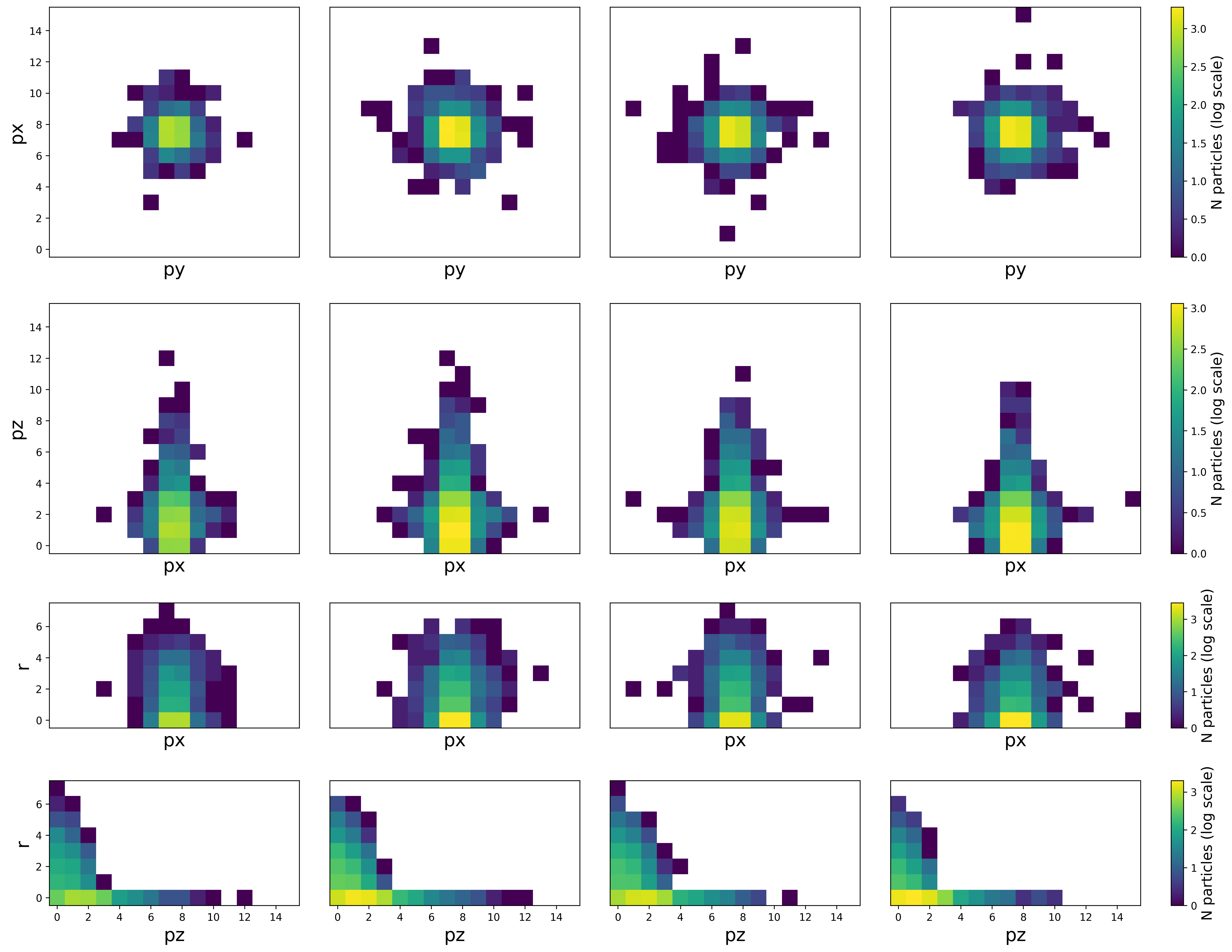}
  \caption{\texttt{CORSIKA} MC-simulated data after applying the adaptive binning scheme. Each panel corresponds to a projection of the four-dimensional data array onto a specific pair of variables, obtained by summing over the remaining dimensions. Different columns correspond to randomly selected events.  Each bin contains the logarithm (base 10) of the number of particles contained within it.}
  \label{fig:MC_simulations_Img}
\end{figure}

Fig.~\ref{fig:MC_simulations_Img} illustrates the effect of the proposed binning scheme on a representative simulated event. Because the encoded tensor has four dimensions, the figure presents two-dimensional projections, obtained by summing over the remaining axes. The colour scales represent the base-10 logarithm of the number of particles contained in each bin.

\section{Generative Adversarial Network model} \label{sec:gan_model}

Generative Adversarial Networks (GANs) constitute a powerful class of generative models that frame data generation as a two-player game between competing neural networks. Within this framework, a \textbf{generator} learns to synthesise data from a latent noise distribution, while a \textbf{discriminator} strives to distinguish real samples, such as those obtained from \texttt{CORSIKA} simulations, from generated ones. GANs have demonstrated remarkable capability in producing highly realistic and visually convincing data across a wide range of domains. However, their training remains notoriously challenging. Recent advances~\cite{Yazdani:2025} have introduced a suite of techniques to improve GAN training dynamics. These methods (including Wasserstein losses, gradient penalties,  ensemble strategies, and Self-Attention modules, as reported for example in Refs.~\cite{Arjovsky:2017, Gulrajani:2017, Zhang:2019, Tronchin:2025}) aim to address critical challenges, such as capturing long-range dependencies and global structure in complex data, preventing vanishing or exploding gradients, and mitigating mode collapse. Collectively, these developments have substantially improved the reliability and fidelity of the GAN outputs, enabling more robust, high-quality data synthesis across a range of domains.

Our approach to fast air shower simulation is based on a Wasserstein Generative Adversarial Network with Gradient Penalty (WGAN-GP)~\cite{Goodfellow:2014, Arjovsky:2017, Gulrajani:2017}, enhanced with Self-Attention~\cite{Zhang:2019}, to better capture the global correlations and complex dependencies inherent in multi-dimensional air-shower data. Compared to the original Generative Adversarial Network (GAN) formulation~\cite{Goodfellow:2014}, which minimizes the Jensen--Shannon divergence between the model and data distributions, the Wasserstein formulation introduces the Earth Mover’s (Wasserstein-1) distance as a smoother and more informative loss function. This modification yields a continuous and differentiable objective that correlates with the quality of generated samples and substantially mitigates issues such as vanishing gradients and mode collapse~\cite{Arjovsky:2017}. This framework was selected for its ability to provide more stable and efficient adversarial training, which is particularly useful when dealing with complex, multi-dimensional distributions typical of particle shower data. 

The inclusion of the gradient penalty term~\cite{Gulrajani:2017} enforces a Lipschitz continuity constraint on the discriminator in a soft and data-dependent manner, replacing the crude weight-clipping approach of the original WGAN. This regularisation ensures more reliable gradient flow to the generator, and provides a way to prevent the discriminator from overfitting to local features. As a result, the WGAN-GP framework provides a more robust learning signal throughout training, making it well suited for modelling physically complex distributions.

Both the generator and discriminator networks employ convolutional architectures to capture hierarchical and spatially localised patterns in the data. To further enhance global coherence and represent long-range dependencies that are often difficult to model using purely convolutional layers, Self-Attention mechanisms~\cite{Zhang:2019,Vaswani:2017} are integrated into both networks. The Self-Attention modules allow the model to dynamically weight feature interactions across distant spatial regions, which is particularly beneficial for complex data with intricate structures (such as high-resolution images or multi-dimensional data like air showers), enabling it to synthesise globally consistent and structurally correlated outputs. This combination of convolutional and Self-Attention operations have been shown to improve the overall quality of generated samples, especially in high-dimensional domains~\cite{Zhang:2019, Brock:2018}, providing the foundation for fast and accurate air shower simulations.

\subsection{Adversarial networks specifications}

The models employed in this study have been implemented in \texttt{TensorFlow/Keras} version 2.17.0 and \texttt{Python} version 3.12.3. Training was conducted for \num{4000} epochs on a system equipped with an NVIDIA~A40 GPU (48~GB of memory). The training process lasted approximately 74 hours.

\begin{figure}[hbt]
  \centering
  \includegraphics[width=\textwidth]{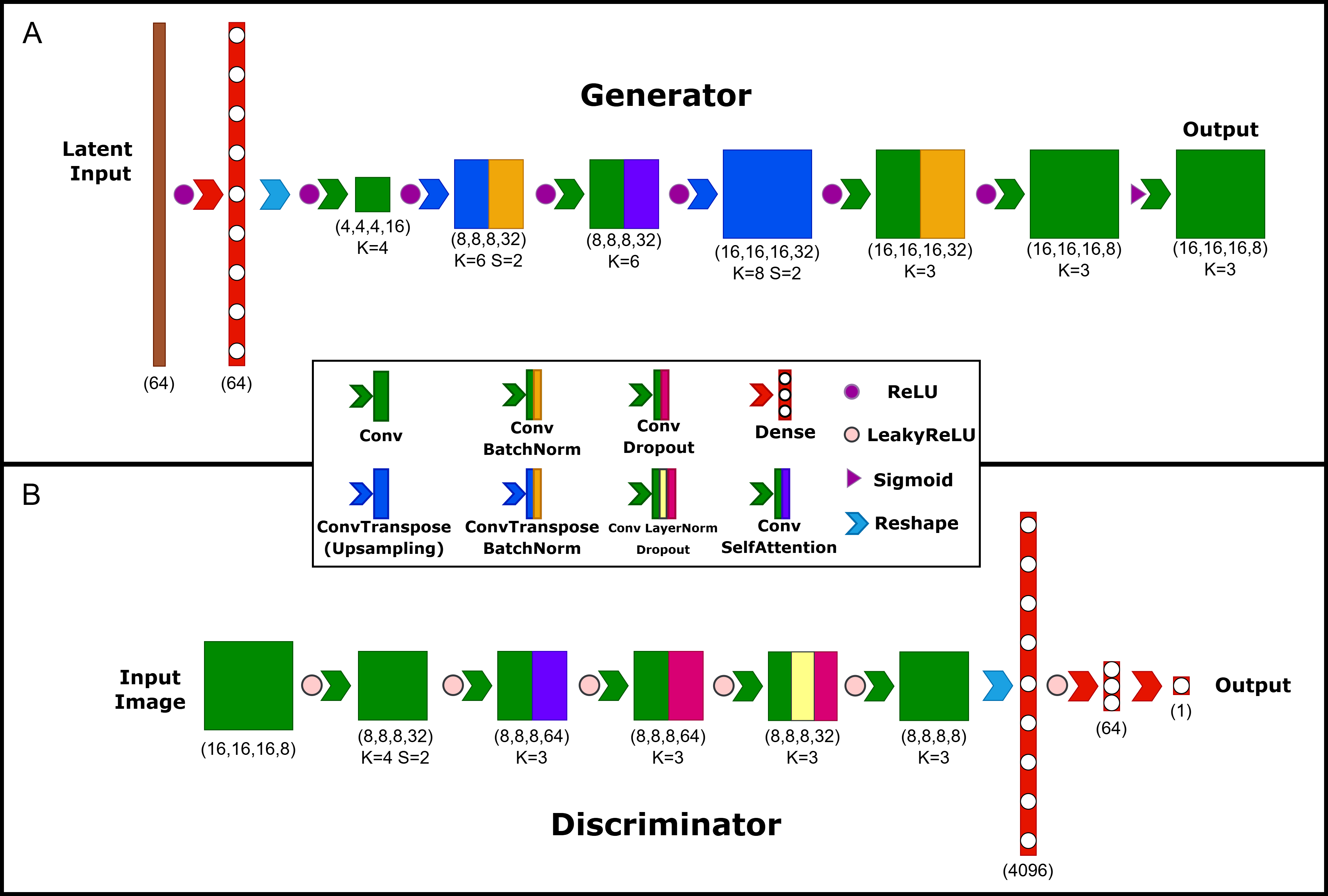}
  \caption{Adversarial Networks architectures. Panel A shows the generator network. Panel B shows the discriminator. The central inset represents a legend for the different layers and activation functions employed. \textit{ConvSelfAttention}, \textit{ConvDropout}, \textit{ConvLayerNorm}, and \textit{ConvBatchNorm} indicate convolution, followed by a \textit{Self-Attention}, \textit{Dropout}, \textit{Layer Normalization}, or \textit{Batch Normalization} operation, respectively, as described in the text. Numbers in parentheses under each layer indicate its shape. The values of $K$ under each convolutional layer indicate the corresponding kernel size (equal for each dimension). The values of $S$ indicate, when present, the corresponding stride employed at that layer. The \textit{Reshape} procedure (light blue arrow) only reshapes a 4D previous layer in a one-dimensional array, or vice versa, without affecting any value.} 
  \label{fig:Architectures}
\end{figure}

The WGAN was trained using a batch size of 512, the largest possible given our GPU memory constraints, as larger batch sizes have been shown to improve gradient stability and feedback quality during adversarial training~\cite{Brock:2018}. The generator receives inputs sampled from a 64-dimensional latent vector, whose components are independently drawn from a standard normal distribution. Figure~\ref{fig:Architectures} shows the architecture of the adversarial networks used here.

The generator receives random latent vectors and maps them in output tensors of size $16 \times 16 \times 16 \times 8$, through a sequence of fully connected layers, 3D convolutional layers, 3D transposed convolutional layers, and \texttt{ReLU} activations~\cite{Krizhevsky:2012}. A final sigmoid activation function~\cite{Han:1995} constrains the output tensor to the range $[0,1]$, matching the normalisation of the training data. A Self-Attention module~\cite{Zhang:2019} is included at the intermediate feature level to capture non-local spatial dependencies and improve the global coherence of the generated volumetric data. 

The discriminator network receives as input arrays of size $16 \times 16 \times 16 \times 8$ and outputs a scalar which represents the discriminator’s estimate of the Wasserstein distance. The network architecture is divided into two stages. The first is represented by 3D convolutional layers. They provide a very efficient way to extract relevant features from grid-like data~\cite{LeCun:2015}. In this stage, we also employed dropout~\cite{Wan:2013}, useful for preventing overfitting), and layer normalisation~\cite{Ba:2016}, common in GANs discriminators).  A Self-Attention block is also included in the discriminator to enhance sensitivity to global structures. The last stage is composed of fully connected layers, which serve to aggregate and interpret the features extracted by the preceding convolutional structure. Every layer uses a \texttt{LeakyReLU}~\cite{Maas:2013} activation function, except for the final dense layer, which, as usual for WGANs, has no non-linear activation function.

For each generator update, the discriminator was trained over four consecutive iterations to ensure that the Wasserstein distance was well approximated at each step. 

For the generator training, we used the \texttt{Adam} (Adaptive Moment Estimation) optimiser~\cite{Kingma:2014}, a stochastic gradient descent method based on the adaptive estimation of first- and second-order moments. The learning rate was set to $5 \times 10^{-5}$, with parameters $\beta_1 = 0.3$, $\beta_2 = 0.99$, and $\epsilon = 10^{-3}$, where $\beta_1$ and $\beta_2$ are the exponential decay rates for the first and second moment estimates, respectively, and $\epsilon$ is a small constant added to prevent division by zero.

The discriminator has been optimised by \texttt{RMSprop} (Root Mean Square Propagation)~\cite{Hinton:2012}, an adaptive learning rate algorithm that adjusts the step size for each parameter based on the moving average of the squared gradients. The optimizer was configured with a learning rate of $2 \times 10^{-4}$, the momentum parameters set to $0.3$, and $\epsilon = 10^{-3}$. 

This two-timescale learning rate setup (\texttt{TTUR}) has been shown to facilitate a stable balance between the generator and discriminator updates~\cite{Heusel:2017}. The gradient penalty coefficient was fixed at $\lambda_{\mathrm{GP}} = 10$, following the prescription of Ref.~\cite{Gulrajani:2017}.

\section{Results}

We continuously monitored the training process by saving model checkpoints every 30 epochs. As reported in the literature, GANs are known to be challenging to train: even under carefully tuned conditions and optimal training practices, GANs often display unstable dynamics, such as non-convergent oscillations or mode collapse, in which the generator reproduces only a limited subset of the training data~\cite{Brock:2018, Che:2016, Metz:2016}. 

To better monitor these effects, investigate the evolution of the model, and assess the physical relevance of the generated data during training, we evaluated the performance of the generator at each checkpoint by producing 2048 synthetic events. The resulting distributions of the total number of particles per event were then compared with those obtained from the data derived from MC simulations. While the discriminator loss converges steadily around epoch 500, the monitored distribution does not stabilise, oscillating rather than converging to a limiting distribution.

\begin{figure}[hbt]
  \centering
  \includegraphics[width=0.8\textwidth]{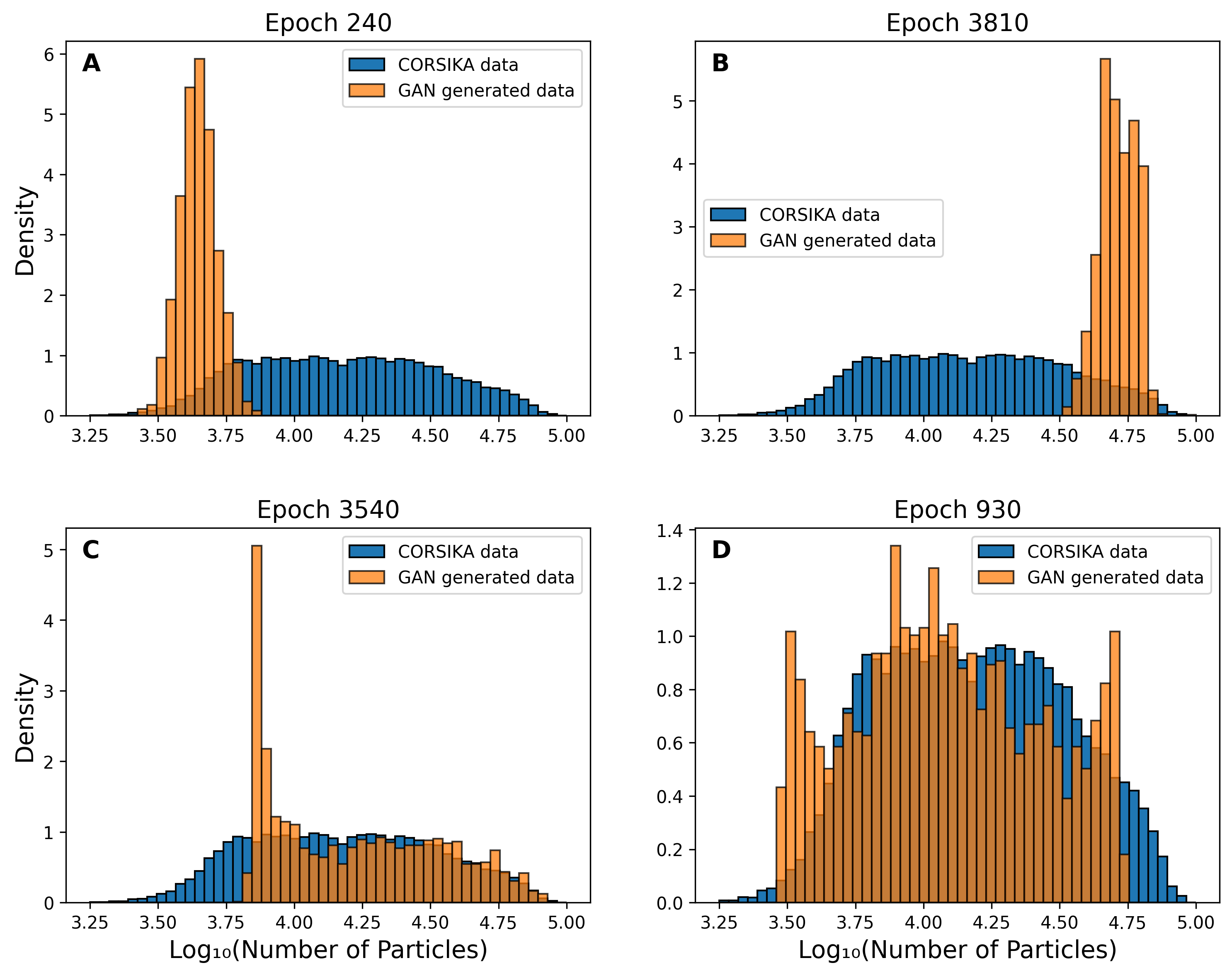}
  \caption{Distributions of the total number of particles per event at some representative epochs, normalised to unit area. Orange bars refer to different fixed-epoch WGAN-generated data, while blue bars refer to the \texttt{CORSIKA} training data. The x-axes show the base-10 logarithm of the number of particles. Each panel refers to a different epoch.} 
  \label{fig:Npart_epoche}
\end{figure}

As shown in Fig.~\ref{fig:Npart_epoche}, we observed the generator to exhibit variable behaviour across training epochs. At certain checkpoints, it was able to reproduce only a subset of the training data distribution (Panels A and B), while at others it captured most modes but did not populate specific regions, such as the leftmost region in Panel C. Such behaviour reflects partial mode coverage, where each epoch tends to emphasise distinct modes of the training distribution. In some cases, the generated distributions appear broadly consistent with the training one yet remained slightly irregular, as illustrated in Panel~D. The alternation between well-shaped and incomplete distributions recurs intermittently throughout the entire training and persists over time, indicating a lack of convergence. In such a scenario, selecting a single ``best'' epoch is challenging.

To mitigate these effects, inspired by studies demonstrating that aggregating multiple WGANs can enhance mode coverage and achieve a better approximation of the training distribution~\cite{Tronchin:2025, Eilertsen:2021, Zhang:2019b}, we adopted an ensemble strategy. Specifically, since we noticed that models corresponding to different epochs of training showed a tendency to cover the regions of the monitored distribution in different ways, we constructed our ensemble by combining generator models saved at different epochs. This approach, known as the self-ensemble strategy~\cite{Wang:2016}, offers the advantage of requiring only a single training, making it computationally efficient, while still benefiting from the diversity of representations learned over time.

\begin{figure}[hbt]
  \centering
  \includegraphics[width=0.6\textwidth]{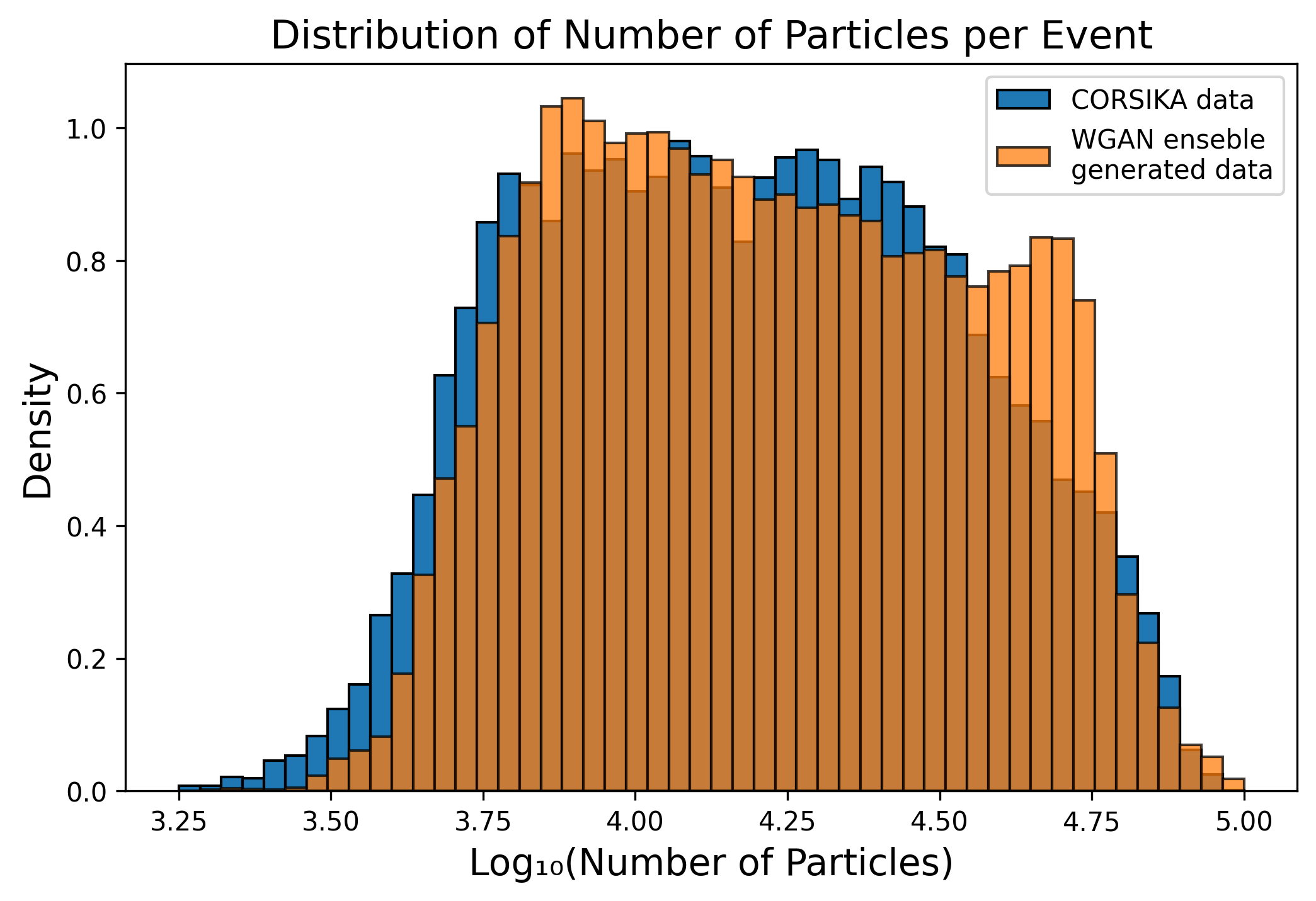}
  \caption{Distributions of the total number of particles per event, normalised to unit area. Orange bars refer to data generated by the WGAN ensemble (\num{2048} events for each of the 57 selected epochs). Blue bars indicate the corresponding distribution using the \texttt{CORSIKA} training data. The x-axis reports the base-10 logarithm of the total particle count per event.} 
  \label{fig:Npart_all}
\end{figure}

To build our ensemble, we selected the best generators through the epochs, using the distribution of the number of particles as selection criterion. Specifically, we computed the Wasserstein distance between the training distribution and those derived from generated data at fixed epochs, and selected all generators whose distance was less than or equal to 0.1. To provide context for this threshold, we note that a Wasserstein distance of 0.1 corresponds to the distance between two Gaussian distributions with unit variance, whose means differ by 0.1~\cite{Givens:1984}. In addition, we considered only generators from epochs greater than 500, since, as we observed earlier, this is the point at which the discriminator reaches equilibrium. This selection resulted in an ensemble of 57 WGANs. Figure~\ref{fig:Npart_all} shows the distribution of particles per event obtained by aggregating the ensemble, with each WGAN generating 2048 events. The aggregated distribution produced by the ensemble exhibits a smoother and more regular behaviour than those generated by single networks. Moreover, it provides a visibly closer agreement with the training data distribution. The Wasserstein distance between the distribution derived from the ensemble and the one derived from \texttt{CORSIKA} data is 0.04, and it quantitatively indicates a significantly improved match relative to any single WGAN.

\subsection{Generated data inspection}

As discussed in Section~\ref{sec:gan_model}, the training procedure required approximately 74 hours for a total of \num{4000} epochs. This represents the only computationally-demanding phase of the workflow, as data generation with the trained model is extremely fast. Once trained, the GAN is capable of producing \num{30000} samples (in the form of four-dimensional binned tensors; see Sec.~\ref{sec:Data_preprocess}) in under one minute.

\begin{figure}[hbt]
  \centering
  \includegraphics[width=\textwidth]{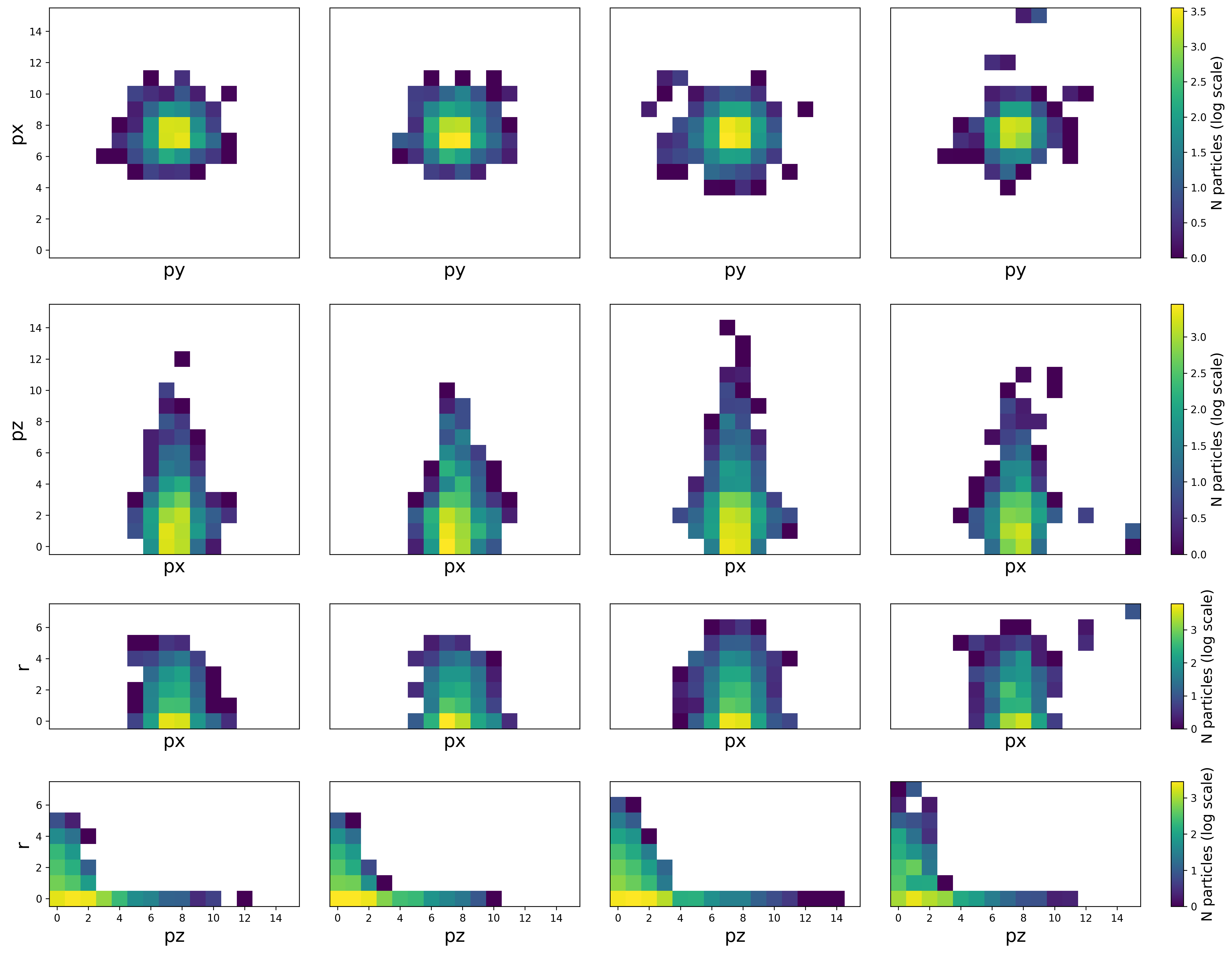}
  \caption{WGAN ensemble generated data. Different columns correspond to different randomly-selected generated events. Each panel corresponds to a projection of the four-dimensional generated array onto a specific pair of variables, obtained by summing over the remaining dimensions.  Each bin contains the logarithm (base 10) of the number of particles contained within it. For comparison, refer to Fig.~\ref{fig:MC_simulations_Img}, which shows the analogous projection from some representative \texttt{CORSIKA}-simulated events.}
  \label{fig:Generated_Data_Img}
\end{figure}

Figure~\ref{fig:Generated_Data_Img} displays some representative samples generated by the WGAN ensemble. Qualitatively, a good agreement is observed between the generated samples and the images derived from the MC-simulated training data (see Fig.~\ref{fig:MC_simulations_Img}). This visual consistency holds across all pairs of binned phase-space variables, indicating that the WGANs successfully reproduce the global structures and relative intensity patterns present in the original data. 

\begin{figure}[hbt]
  \centering
  \includegraphics[width=0.8\textwidth]{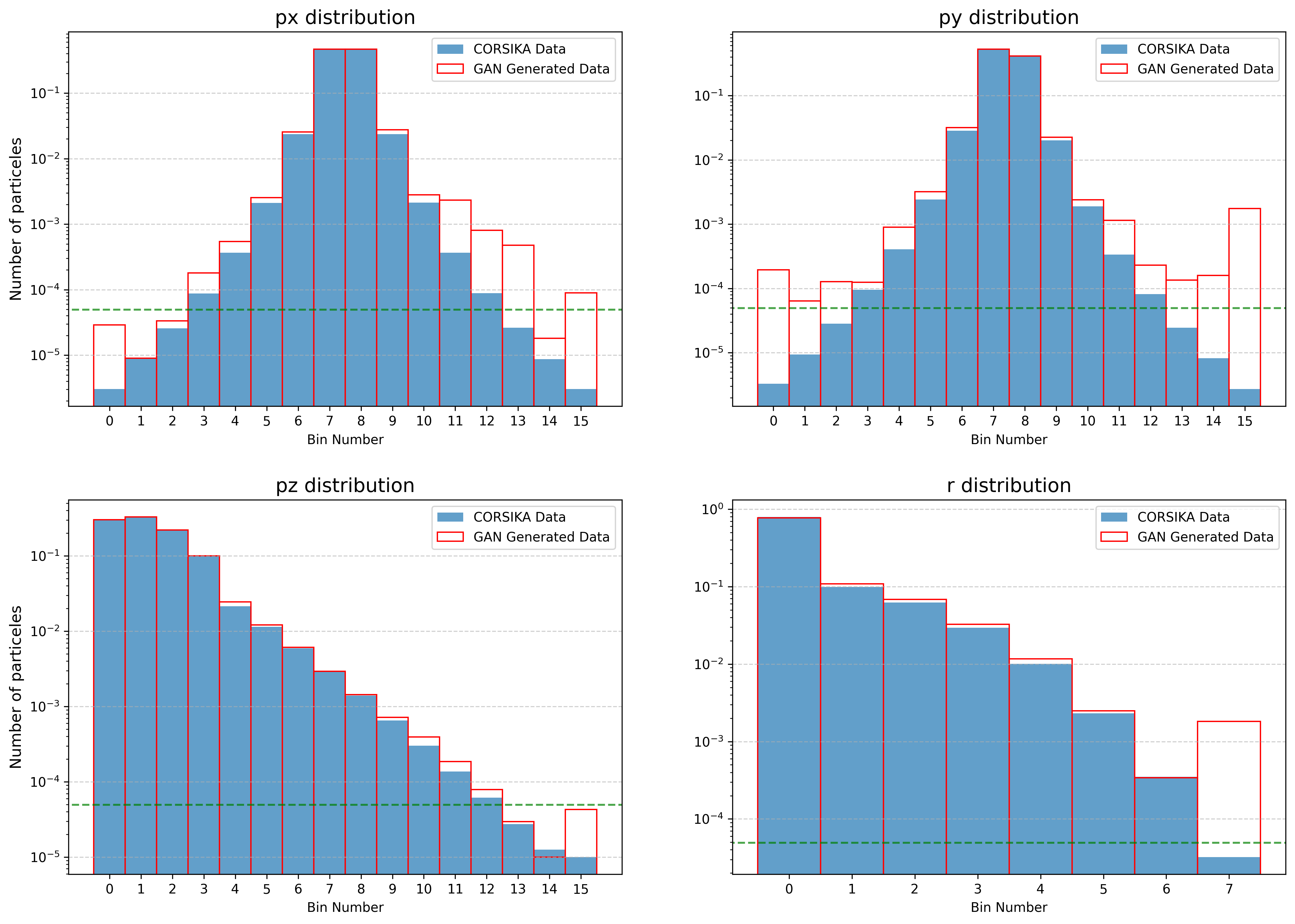}
  \caption{Normalised one-dimensional bin-occupancy distributions for each dimension in data tensors (logscale). Each distribution is scaled to unit area. Light-blue bars refer to \texttt{CORSIKA} MC-simulated data, while red bars refer to WGAN-generated data. Green dashed lines indicate in each panel the level corresponding to a bin filled with 1 particle.}
  \label{fig:Final_Distributions}
\end{figure}

Figure~\ref{fig:Final_Distributions} presents the one-dimensional bin occupancy distributions for each dimension of data tensors. For every examined quantity, the corresponding plot compares the bin-occupancy distribution between the WGAN-ensemble-generated data and the \texttt{CORSIKA} training dataset. A dashed horizontal green line marks the reference level corresponding to an average bin occupancy of one particle per event. Focusing on bins where the \texttt{CORSIKA} data exceeds this reference level, the WGAN-generated distributions exhibit desired trends: for the $p_x$ and $p_y$ components, occupancy decreases moving toward the extremal bins; for the $p_z$ and $r$ components, occupancy decreases toward the higher bin indices. The most noticeable discrepancies arise in bins with low occupancy, where the generated dataset tends to overpopulate these bins compared to the \texttt{CORSIKA} data. Among these distributions, we notice the $p_z$ and $r$ ones to be reproduced with higher fidelity.

\begin{figure}[hbt]
  \centering
  \includegraphics[width=0.8\textwidth]{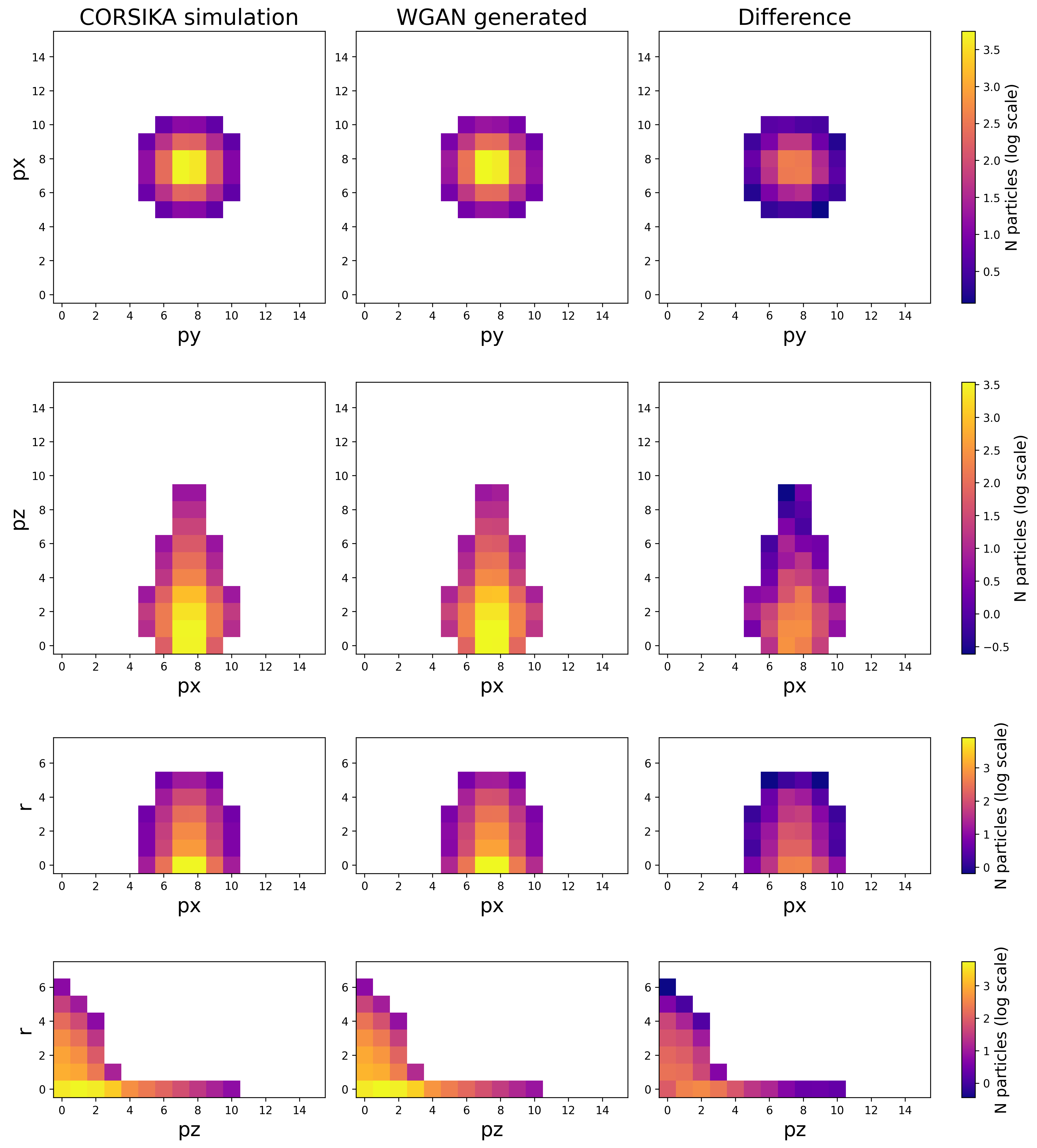}
  \caption{Mean event arrays projected onto pairs of dimensions. The first column indicates the mean computed using the \texttt{CORSIKA} MC-simulated data, the second refers to the WGAN-ensemble generated data, and the last presents their difference. The differences are found to be approximately one order of magnitude smaller than the absolute values. Only bins with a mean occupancy of at least five particles in the \texttt{CORSIKA} data are included to ensure statistical reliability.}
  \label{fig:Mean_image}
\end{figure}

While individual event comparison (see figures \ref{fig:MC_simulations_Img} and \ref{fig:Generated_Data_Img}) offers valuable qualitative insight into the visual fidelity of the generated data, a more comprehensive view is needed. To this end, we computed the mean event tensor for both the \texttt{CORSIKA} training data and the WGAN-ensemble--generated samples. Each mean image shown in Fig.~\ref{fig:Mean_image} represents the average occupancy of particles across all events, projected onto pairs of phase-space dimensions. This approach allows for a direct, large-scale comparison of the distributions reproduced by the generative model. Here we consider only bins with a mean occupancy of at least five particles, as those with lower statistics showed significant fluctuations (for example, symmetric regions such as $p_x$ and $-p_x$ displayed noticeable differences in particle counts). The panels display the mean images derived from the \texttt{CORSIKA} simulations and their WGAN-ensemble–generated counterparts, together with the corresponding difference maps. Across all projections, the mean structures of the generated data closely follow those of the \texttt{CORSIKA} reference dataset. Overall, the WGAN ensemble successfully reproduces the global morphology of the showers, with discrepancies typically an order of magnitude smaller than the mean signal.

\section{Conclusions} \label{sec:concl}

In this work, we present \texttt{GAIAS2} (Generative Artificial Intelligence for Air Shower Simulation), an application of Generative Adversarial Networks (GANs) to the fast simulation of extensive air showers induced by primary cosmic rays in the atmosphere. The model was trained on a large, high-quality dataset of proton-induced extensive air showers generated with the \texttt{CORSIKA} Monte Carlo program. \texttt{GAIAS2} successfully reproduces key features of the underlying particle distributions, including the energy and spatial spectra of muons at ground level.

The WGAN-GP architecture, enhanced with Self-Attention mechanisms, was capable of learning the complex multi-dimensional structure of the air-shower phase space, producing physically consistent particle distributions without the need for explicit parametric modelling. An ensemble strategy was adopted to mitigate partial mode coverage, significantly improving the stability of the generated data and ensuring a more complete representation of the training distribution. Quantitatively, the ensemble of 57 networks achieved a Wasserstein distance of 0.04 with respect to the training data distribution, marking a substantial improvement compared to single-network generation.

Once trained, the \texttt{GAIAS2} model enables a substantial acceleration of the simulation process: the generation of $3\times10^4$ showers requires less than one minute on a single GPU, corresponding to a speed-up factor of approximately $\mathcal{O}(10^{\text{4}})$ relative to full Monte Carlo simulations with \texttt{CORSIKA} under comparable conditions. This result highlights the potential of generative modeling to complement, or partially replace, traditional Monte Carlo pipelines, leading to significant reductions in computational time and energy consumption for large-scale astroparticle physics studies.

This approach is independent of any specific detector geometry or response model. Unlike approaches designed to emulate detector-level observables, \texttt{GAIAS2} is trained directly on the underlying physical phenomenon allowing it to remain general and easily adaptable to different experimental configurations or subsequent detector-response simulations.

Future developments will focus on extending the present framework in several directions. First, we plan to generalise the model to simultaneously generate multiple particle species (e.g., muons, electrons, photons, and hadrons) within the same event, capturing their mutual correlations and shared dependencies in the air-shower evolution. Second, the training dataset will be expanded to include a broader range of primary energies, inclinations, and particle types, thereby improving the generalisation capability of the model. Finally, coupling the generative model with differentiable or hybrid simulation pipelines could enable real-time conditioning on shower parameters and adaptive sampling across the full cosmic-ray spectrum.

The results presented here demonstrate that generative artificial intelligence can provide a viable and efficient path towards fast, high-fidelity air-shower simulations, with potential applications ranging from cosmic-ray composition studies to the design and optimisation of next-generation astroparticle detectors.

\section*{Software and Data Availability}

The software described in this article is released under the MIT License.  
It may be freely used, modified, and distributed, provided that the conditions of the license are respected.  
Derivative works and commercial use are permitted.  
The full text of the MIT License can be found at \url{https://opensource.org/licenses/MIT}.

\section*{Acknowledgments}

This work has been carried out for the GAIAS2 project - \emph{Generative Artificial Intelligence for Air Shower Simulation - CUP  I57G21000110007} -,funded by a cascade grant from Spoke2 / Istituto Nazionale di Fisica Nucleare, as part of the project ICSC – Centro Nazionale di Ricerca in High Performance Computing, Big Data and Quantum Computing, funded by European Union – NextGenerationEU.

\end{document}